\documentclass{raa}            % referee version: for submission [referee]{raa}
\usepackage{graphicx,times}             %for PS/EPS graphics inclusion, new
\usepackage{natbib}
\usepackage{amssymb,amsmath}
\bibpunct{(}{)}{;}{a}{}{,}

\usepackage[a4paper=true,dvipdfm=true,pagebackref=true]{hyperref}
\hypersetup{colorlinks = true, linkcolor = green, anchorcolor = red, citecolor = blue, filecolor = red, pagecolor = red, urlcolor = red}

\begin{document}

   \title{The AIMS Site Survey
\,$^*$
\footnotetext{$*$ Supported by the National Natural Science Foundation of China.}
}
%   \subtitle{I. Place Your Subtitle Here}

   \volnopage{Vol.0 (20xx) No.0, 000--000}      %%preserved for Editor. DOn't remove!
   \setcounter{page}{1}          %%starting page, preserved for Editor. DOn't remove!

   \author{Xing-Ming Bao
      \inst{1}
   \and Jian Wang
      \inst{1}
   \and Shuai Jing
      \inst{1}
   \and Yuan-yong Deng
      \inst{1,2}
   \and Dong-guang Wang
      \inst{1,2}
   \institute{Key Laboratory of Solar Activity, National Astronomical Observatories, Chinese Academy of Sciences, 20 Datun Road, Beijing 100101, China; {\it xbao@bao.ac.cn}\\
     \and
          University of Chinese Academy of Sciences,
          Beijing 100049, China}\\
  %      \and
  \vs\no
   {\small Received~~March 9, 2023 ; accepted~~August 2, 2023~~}}
\abstract{ This paper reports site survey results for the Infrared System for the
Accurate Measurement of Solar Magnetic Field, especially in Saishiteng Mountain, Qinghai, China. Since 2017, we have installed weather station, spectrometer for precipitable water vapor (PWV) and S-DIMM  and carried out observation on weather elements, precipitable water vapor and daytime seeing condition for more than one year in almost all candidates. At Mt. Saishiteng, the median value of daytime precipitable water vapor  is 5.25 mm and its median value in winter season is 2.1 mm. The  median value of  Fried parameter of daytime seeing observation at Saishiteng Mountain is 3.42 cm. Its solar direct radiation data shows that solar average observable time is 446 minutes per day and premium time is 401 minutes per day in August 2019.
%$\cdots\cdots$
\keywords{ site testing--- seeing}
}
   \authorrunning{X. Bao, J. Wang, S.  Jing,  Y. Deng   \&  D. Wang  }            %author_head in even pages
   \titlerunning{Site testing result for AIMS }  % title_head in odd pages

   \maketitle
%% The author head (on even pages) and the title head (on odd pages) will be
%% automatically extracted from \author{} and \title{}. Whenever the title is too long,
%% you will be asked to supply a shorter one by inserting either \authorrunning{} or
%% \titlerunning{} before \maketitle. Anyway, you can specify your own heads.
%%
%%
%% Note: In the following text body of your manuscript, please note several differences from
%%       other major journals:
%% (1) \subsection{Please Capitalize the First Letter of Each Notional Word in Subsection Title}
%% (2) Please Capitalize the First Letter of Each Notional Word in all tables' captions
%
%________________________________________________ sections below
%
\section{Requirement for AIMS candidate sites}           %% first-level sections will be auto-capitalized
\label{sect:intro}

\begin{figure}
   \centering
   \includegraphics[width=15cm, angle=0]{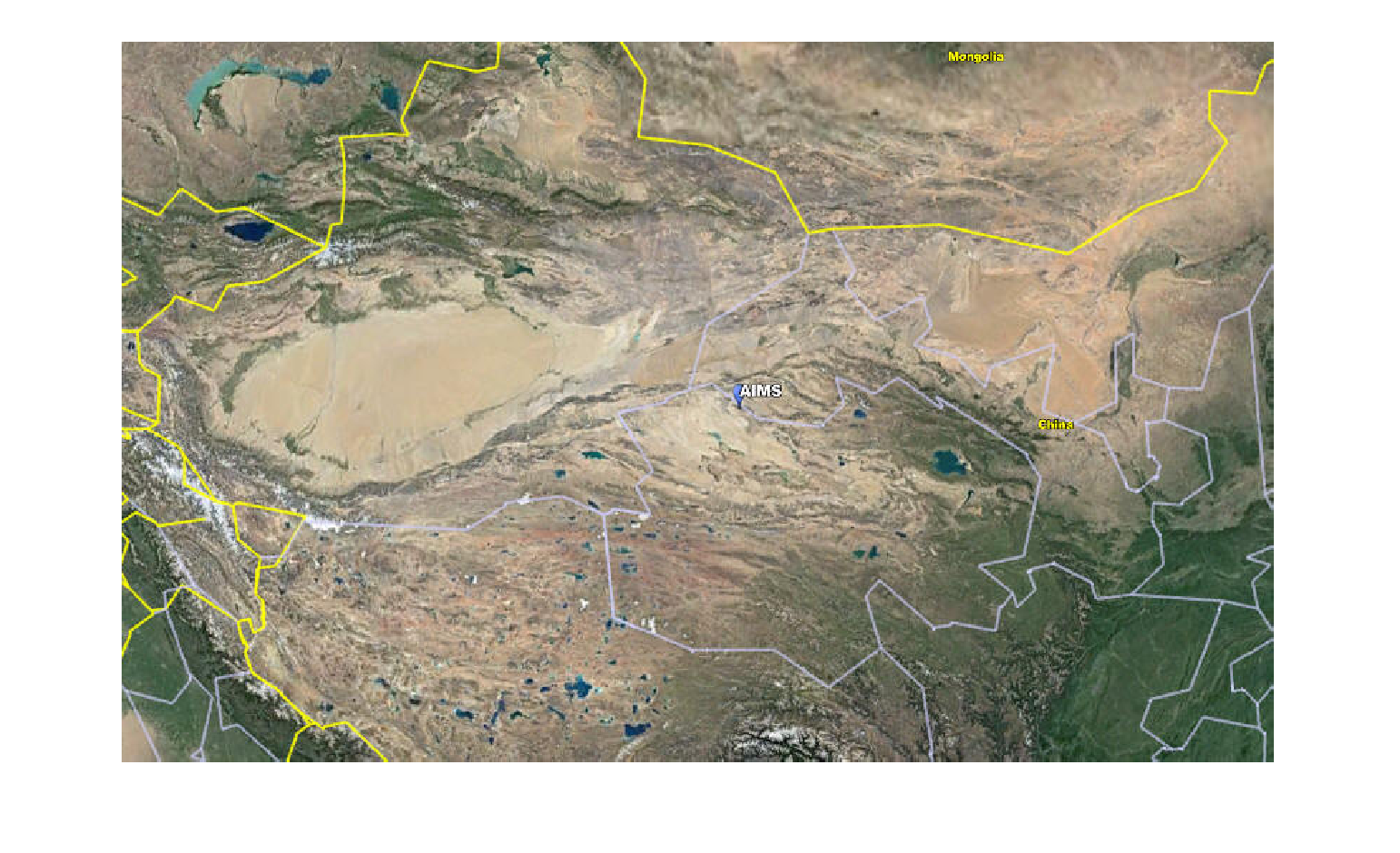}
   \caption{ Mt. Saishiteng is located at north edge of Qiadam Basin, China (N $38^{\circ}$36'45", E 93$^{\circ}$53'45"). }
   \label{Fig:saim}
   \end{figure}

\begin{figure}
   \centering
   \includegraphics[width=10cm, angle=0]{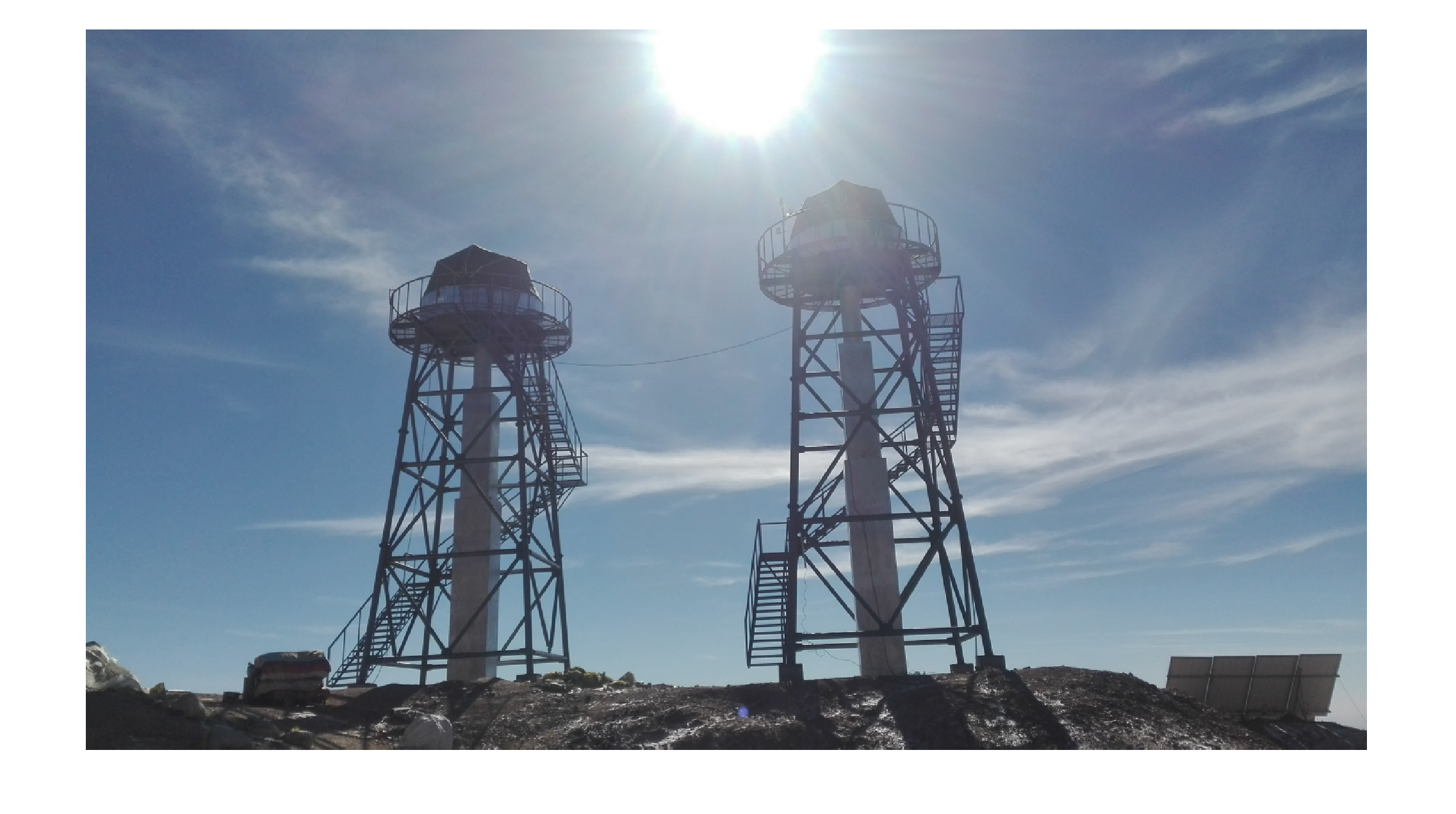}
   \caption{The AIMS site testing tower for monitoring daytime seeing condition (left) with a height of 10 meters. }
   \label{Fig:tower}
   \end{figure}

The Infrared System for the Accurate Measurement of Solar Magnetic Field (AIMS), is a 1 meter telescope  which is dedicated to measure the solar magnetic field at middle infra-red waveband by using a Fourier Transform Spectrometer with high spectral resolution.  In order to maximized performance of AIMS observation, a couple of astronomical environmental factors are considered in site test investigation.  First, the Sun is observed at Mg I 12.3 ${\mu}$m, AIMS requires very low level of perceptible vapor water since atmospheric water vapor content has a strong impact on the transparency of the atmosphere in the infrared and submillimeter domains (\citealt{Kerber+etal+2012}). Second, the daytime  seeing condition need to be measured in order to obtain good daytime image quality which is closely related to AIMS performance. In the nighttime, the seeing condition was usually measured  using Differential Image Motion Monitor (DIMM) such as investigation in Euroupean Southern Observatory at Chile (\citealt{Sarazin+Roddier+1990}), while the daytime seeing condition could be measured by using Solar Differential Image Motion Monitor (S-DIMM) such as in Fuxianhu Lake, China (\citealt{Liu+Beckers+2001}) and in TUG, Turkey (\citealt{Ozisik+Ak+2004}).  Third, AIMS requires as much solar observable time as possible.\\
\\
The site testing investigation for AIMS started in 2016 with two phases. In phase I (2016-2018), we mainly consider the well finished stations with good accommodation and infrastructure conditions due to limited construction period. As a result   Ali in Tibet, Nanshan station in Xinjiang, Delingha station in Qinghai and Daocheng in Sichuan were selected as  first four primary candidate sites. Nanshan and Delingha are well built stations with good accommodation. The altitude of  Ali station is 5100 m above sea level and its logistic is poor and live condition is tough. The site test survey of Daocheng in  daytime has been done more than one year (\citealt{Song+etal+2018}).  The weather station was installed at Nanshan, Ali and Delingha station and a precipitable water vapor spectrometer was deployed in Delingha Station. In phase II (2018-2020), we focus site testing observation on Saishiteng Mountain (38$^{\circ}$36'24"N, $93^{\circ}$53'45"E, with altitude of 4200 m) located in the north edge of Qiadam basin, Qinghai province  (Figure~\ref{Fig:saim}).  It is 50 kilometer east of Lenghu town which is the only inhabited town with an altitude of 2750 m, with arid climate all around. In November, 2018, a S-DIMM was deployed and the daytime seeing condition was carried out more than one year. An observing tower with a height of 10 meters  was built on the Saishiteng mountain (Figure~\ref{Fig:tower}) in November, 2018. In the following sections, we introduce the weather element result in Section 2, show the pwv result in Section 3, describe  the S-DIMM configuration and processing method, display daytime seeing result in Section 4 and discuss the site testing result in section 5.

\section{weather elements}
\label{sect:weather}
\begin{figure}
   \centering
   \includegraphics[width=15cm, angle=0]{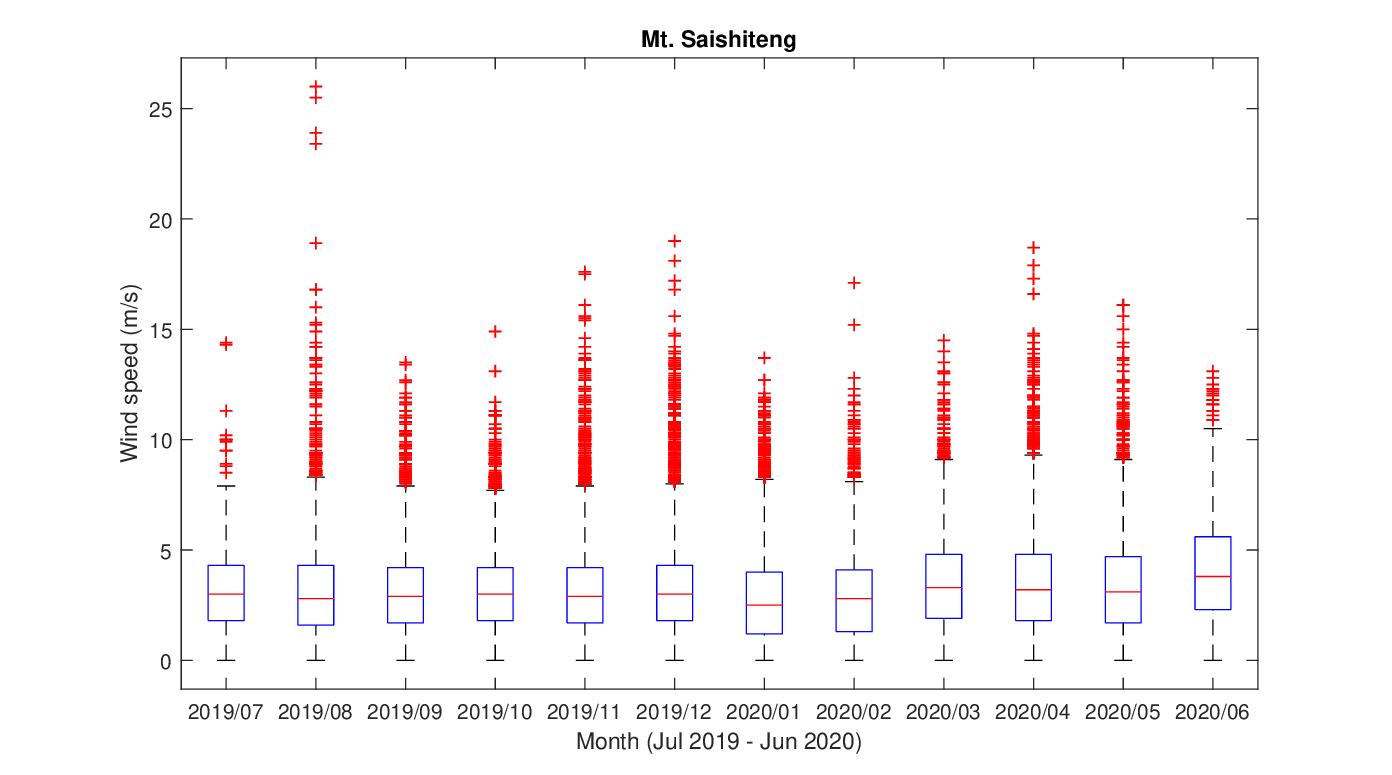}
   \caption{Monthly  variation of wind speed  at Mt. Saishiteng.  The upper tip, upper top of a box, mid-bar in a box, bottom of a box and lower tip represent 95\%, 75\%, 50\%, 25\% and 5\% of the measured data for each month respectively. Plus signs represent the outlier data which are beyond box. }
    \label{Fig:saiwind}
   \end{figure}

\begin{figure}
   \centering
   \includegraphics[width=15cm, angle=0]{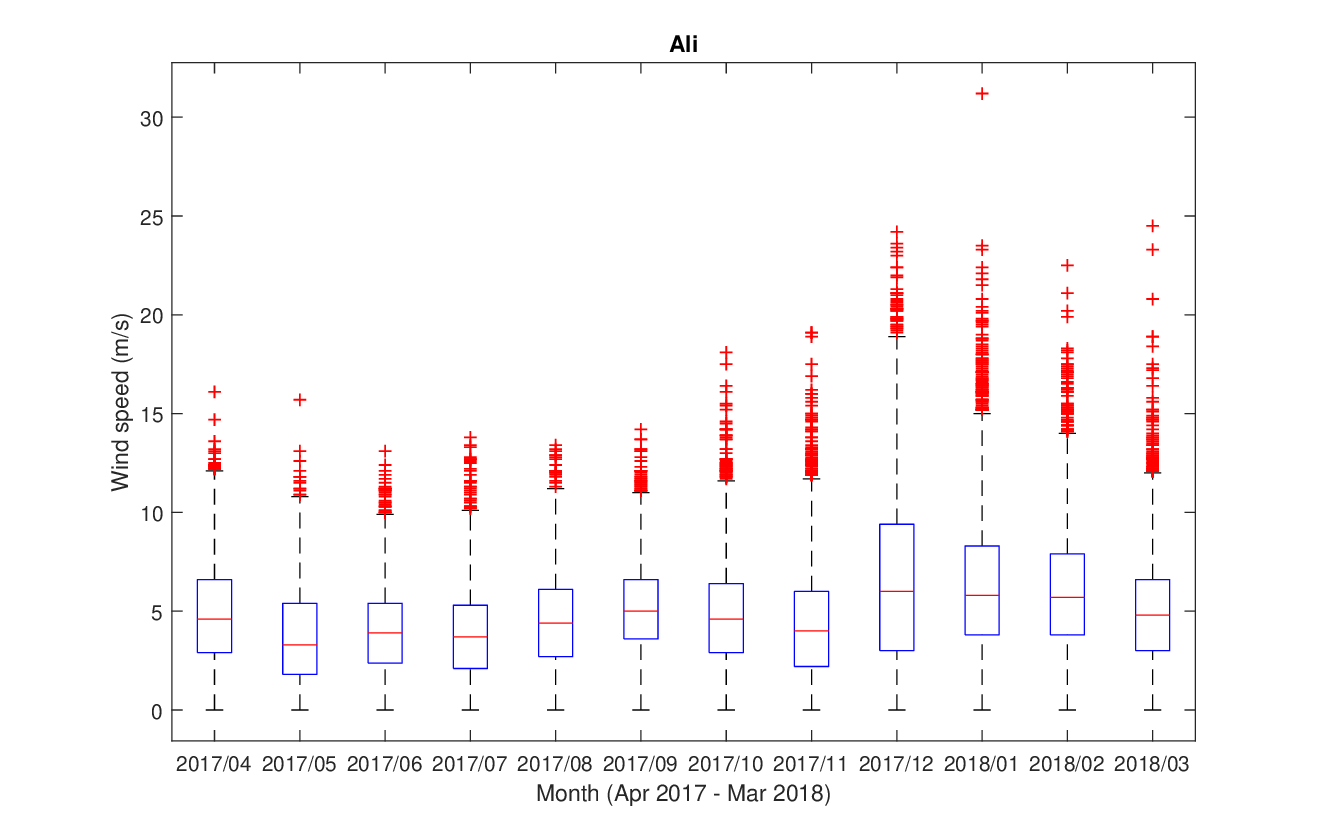}
   \caption{Monthly  variation of wind speed at Ali station. }
   \label{Fig:aliwind}
   \end{figure}

\begin{figure}
   \centering
   \includegraphics[width=15cm, angle=0]{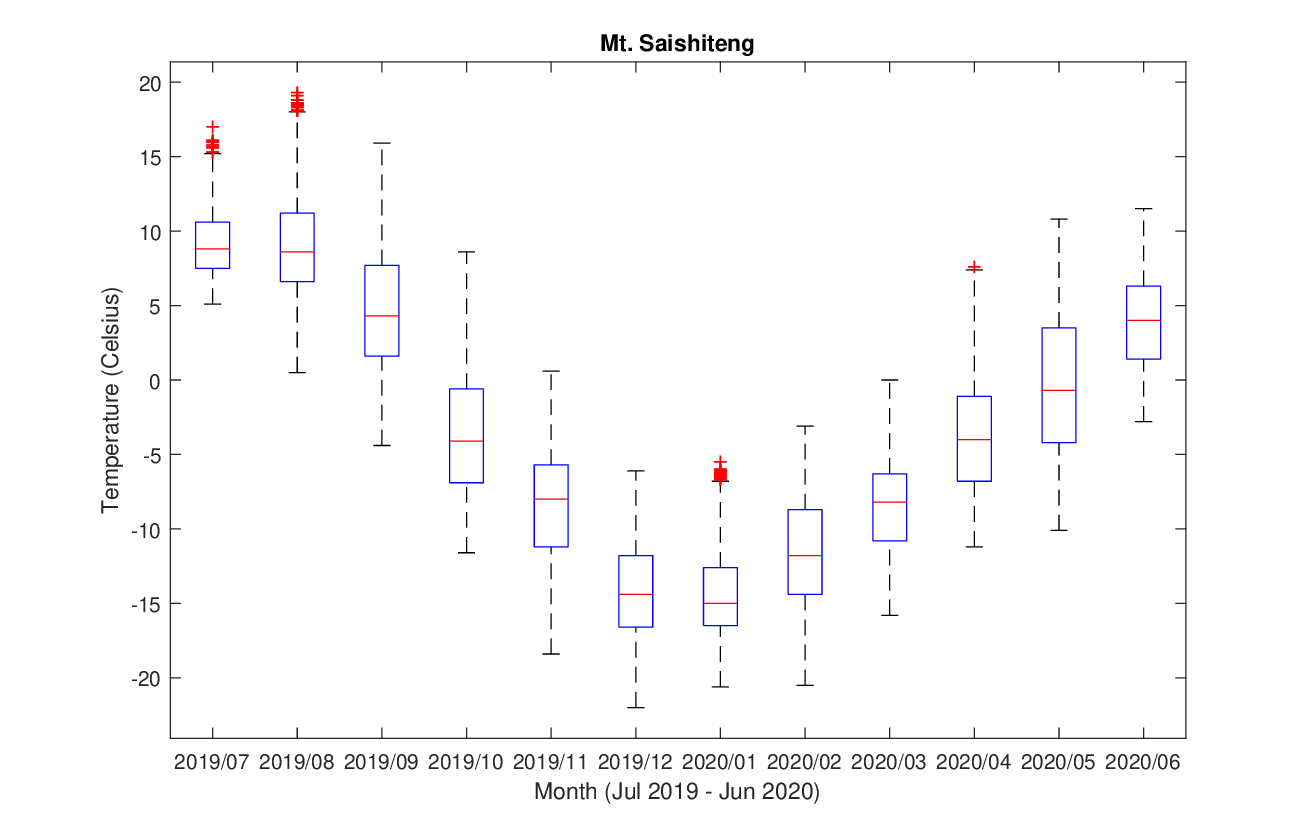}
   \caption{Monthly  variation of temperature at Mt. Saishiteng. The median temperature is -6.5 ${}^{\circ}C$, the highest temperature is 19.2 ${}^{\circ}C$ and the lowest temperature is -22 ${}^{\circ}C$}
   \label{Fig:temp}
   \end{figure}

\begin{figure}
   \centering
   \includegraphics[width=15cm, angle=0]{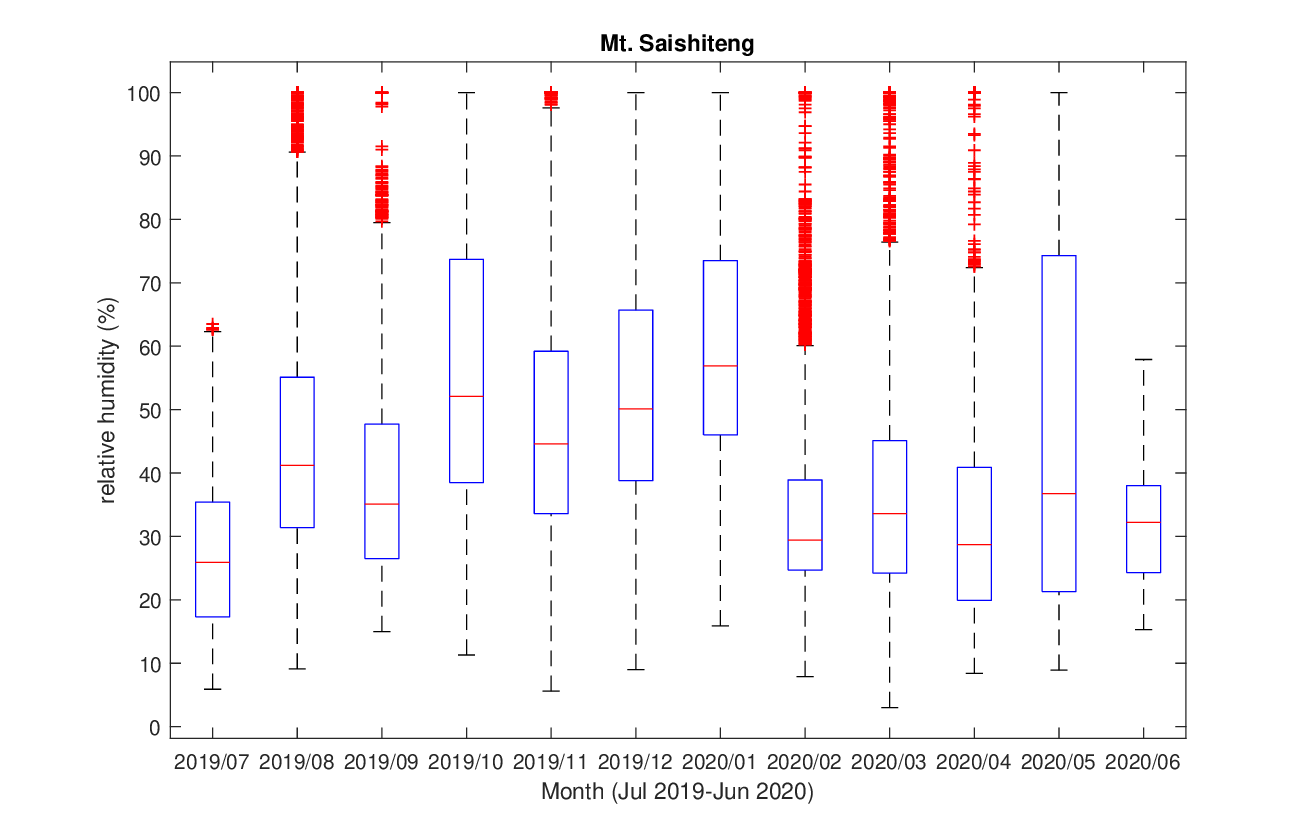}
   \caption{Monthly variation of relative humidity  at Mt. Saishiteng. }
   \label{Fig:humi}
   \end{figure}

\begin{figure}
   \centering
   \includegraphics[width=15cm, angle=0]{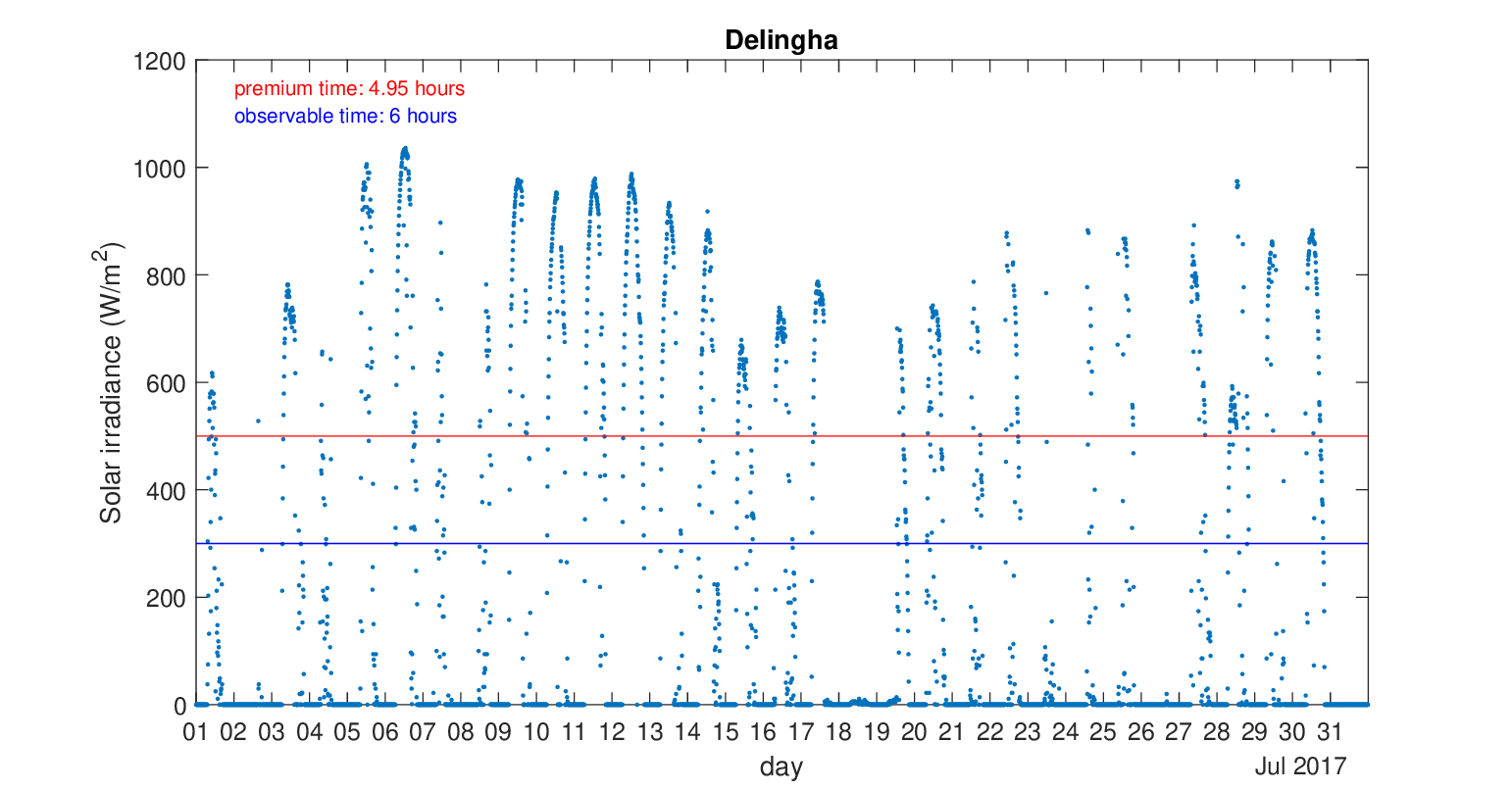}
   \caption{The solar radiance variation  in July, 2017 at Delingha station. Red and blue horizontal lines indicate solar radiation of 500 and 300 W/$m^{2}$ respectively. In each day, its average premium time is   296 minutes (5.95 hours) and  observable time is 360 minutes (6 hours). }
   \label{Fig:derad}
   \end{figure}

\begin{figure}
   \centering
   \includegraphics[width=12cm, angle=0]{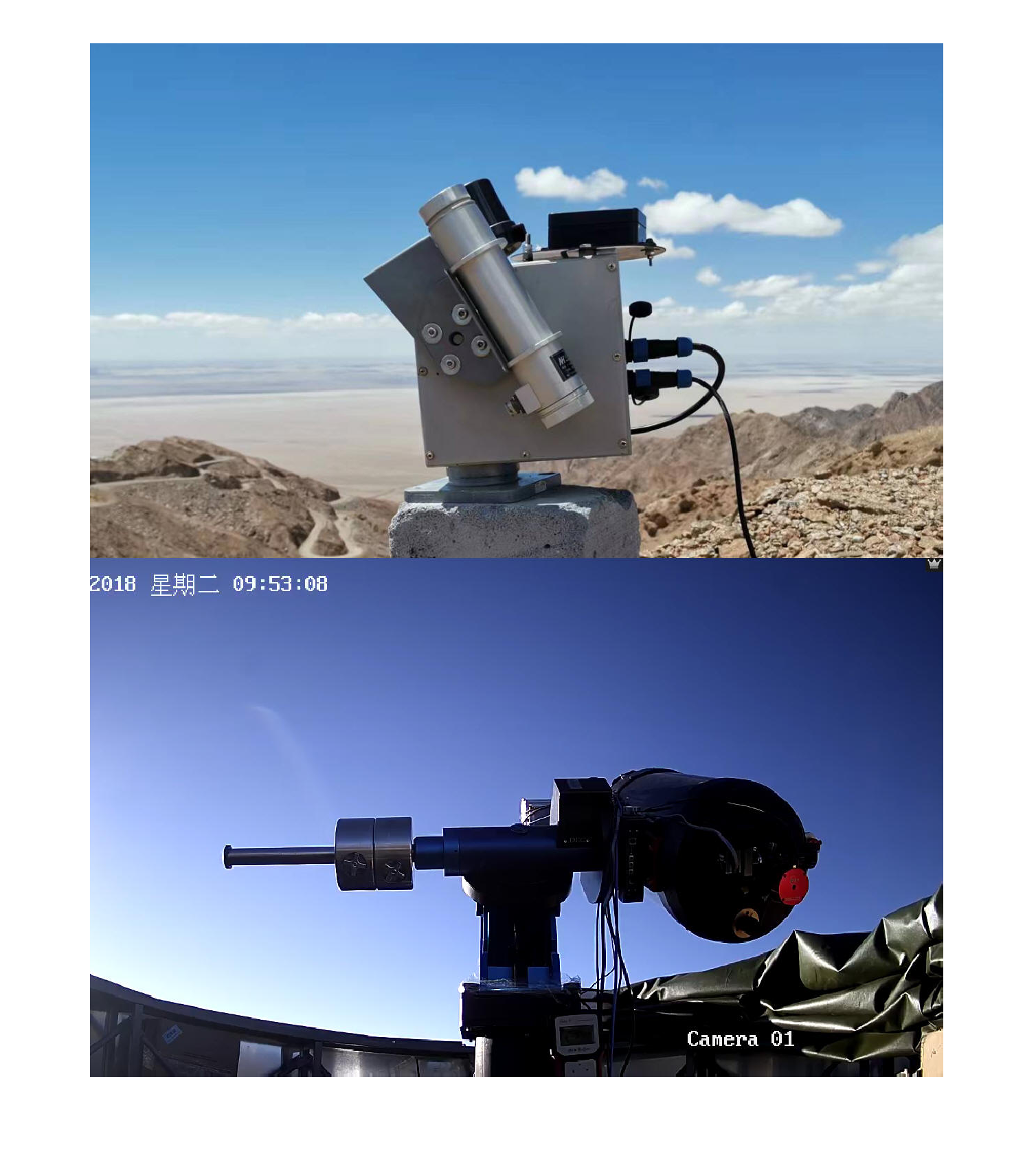}
   \caption{The solar direct radiation meter was mounted on a fully automatic Solar tracker in Mt. Saishiteng (top). The field of view  of solar radiation meter is 5$^{\circ}$\ and  tracking  accuracy of solar tracker is less than 1 $^{\circ}$. S-DIMM and precipitable water vapor spectrometer were mounted together on an ATR  equatorial mounting at Mt. Saishiteng  (bottom).}
   \label{Fig:lhraddimm}
   \end{figure}

\begin{figure}
   \centering
   \includegraphics[width=15cm, angle=0]{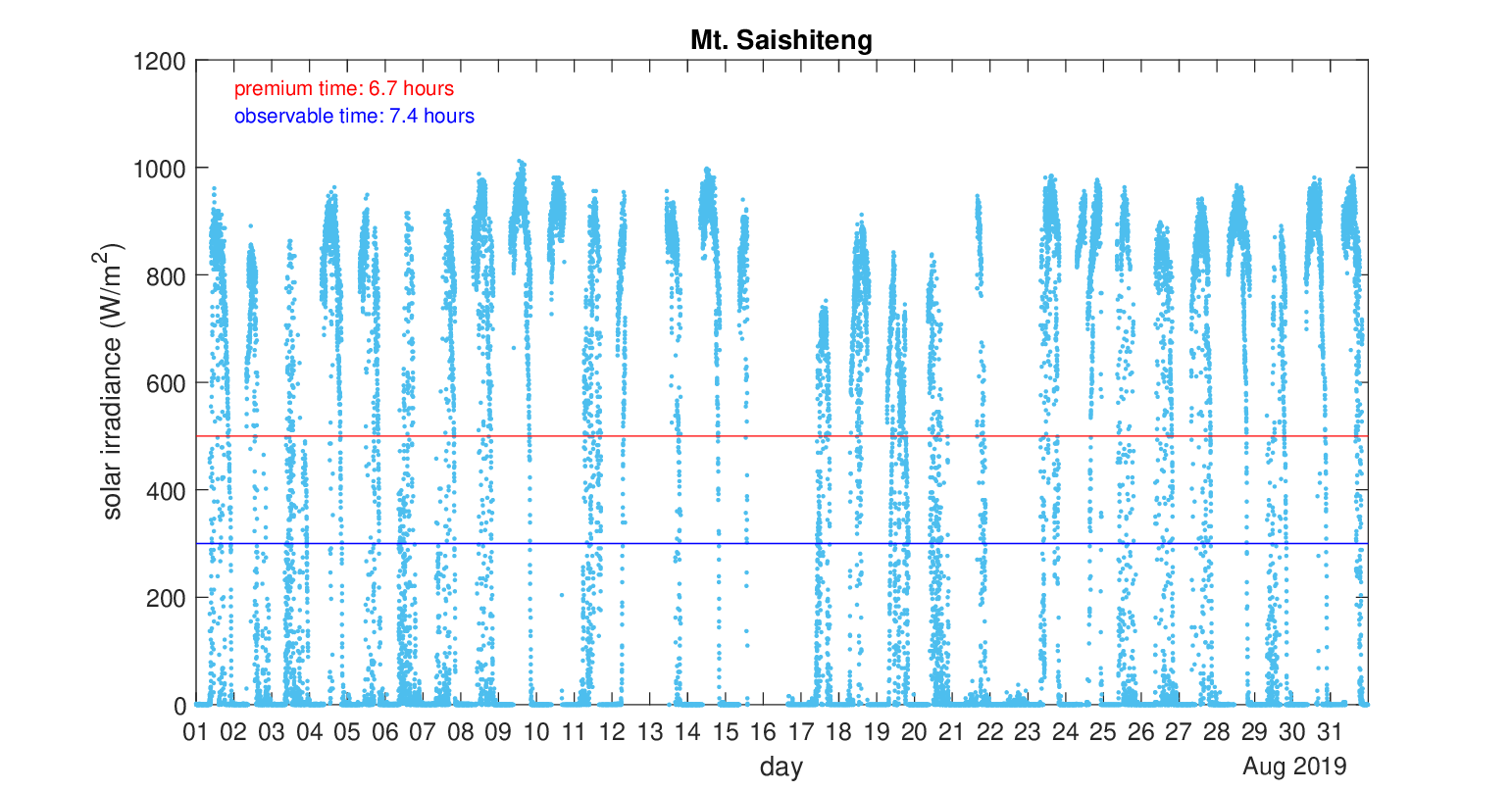}
   \caption{The solar irradiance variation  in August, 2019 at Mt. Saishiteng. Red and blue horizontal lines indicate solar radiation of 500 and 300 W/$m^{2}$ respectively. In each day, its average premium   time is 401 minutes (6.7 hours) and  observable time 446 minutes (7.4 hours). }
   \label{Fig:rads}
   \end{figure}

In Nanshan, Ali,  Delingha and Mt. Saishiteng sites, we have carried out meteorological data observation which includes five weather elements: temperature, relative humidity, wind speed and direction and solar direct radiation. All the weather elements were recorded in the time interval of  1 minute. The wind speed is an important weather element which is related to seeing condition. Figure~\ref{Fig:saiwind} is monthly variation of wind speed in Mt. Saishiteng with median value of 3.2 m/s, and the highest speed of 26.34 m/s.  The wind speed in Ali station is the highest with median value of 4.7 m/s and highest speed of 31.2 m/s (Figure~\ref{Fig:aliwind}).   Table~\ref{Tab:annu-winds} summarizes the wind speed statistics of the four sites, among them, Delingha station has recorded the lowest median value of 1 m/s and the median value in Nanshan station is 2.4 m/s.\\
\\
 Figure~\ref{Fig:temp} shows the monthly variation of temperature in Mt. Saishiteng during July 27, 2019-June 5, 2020 with median value of  -6.5 Celsius, average value of -5.0 Celsius, the highest temperature of 19.2 Celsius and the lowest temperature of -22 Celsius.  Its monthly average temperature is shown in the Table~\ref{Tab:month-temp}. The monthly variation of  relative humidity in Mt. Satshiteng is shown in Figure~\ref{Fig:humi} with median value of 40.3${\%}$  and  average value of 45${\%}$.\\
\\
The total solar irradiance (TSI) is the intensity of solar radiance outside the Earth atmosphere with a constant of around 1361 W/$m^{2}$ (\citealt{Kopp+2021}). Due to the absorption of atmosphere,  the intensity of solar direct radiance is usually lower than 1200 W/$m^{2}$ on the surface of the Earth. On a clear day, the maximum of solar irradiance indicates the absorption (or extinction) of local atmosphere. On a cloudy day, the solar direct radiation reflects the thickness of cloud through the path from observer to the Sun regardless cloud covering the sky outside the Sun. Therefore, we can determine if the Sun can be observed and how long it can be observed from the solar direct radiation data. According our experience, a solar telescope can  observe the Sun continually  when solar direct radiation is more than 500 W/$m^{2}$ and observe the Sun intermittently when solar irradiance ranges between 300-500 W/$m^{2}$ due to cloud or fog. It is hard  to observe the Sun normally when solar direct radiation is below 300 W/$m^{2}$. In this paper, we define solar premium time and observable time as the solar direct radiation is more than 500 W/$m^{2}$ and 300 W/$m^{2}$ respectively.\\
\\
The solar direct radiation meter which is used in AIMS site testing contains a thermoelectric pile which measures the solar irradiance within field of view of $5^{\circ}$ in spectral range of 300-3000 nm. In the early phase,  a semi-automatic tracking system was used to follow the Sun at Nanshan, Ali and Delingha stations. It follows the Sun automatically along right ascension in a speed of 15$^{\circ}$/hour, while it need to be adjusted  manually along declination.  Later we find its performance is poor if no one dedicate to adjust the tube pointing to the Sun once the Sun's declination varies or electricity outage occurs occasionally. Figure~\ref{Fig:derad} are the monthly solar irradiance distribution in June 2017 at Delingha  station with an average premium time of  296 minutes (5.95 hours) and an observable time of 360 minutes (6 hours) per day. The blue and red horizontal lines indicate low limitation of solar radiance 300 W/$m^{2}$ and 500 W/$m^{2}$ for observable hour and premium hour respectively.  In Mt. Saishiteng, we deployed a fully automatic solar tracking system which can follow the Sun in an accuracy less than 1 $^{\circ}$ during the Sun appears above horizon (Figure~\ref{Fig:lhraddimm} top). Figure~\ref{Fig:rads} shows the monthly solar irradiance variation at Mt. Saishiteng in August, 2019. It indicates that the average premium time  is 401 minutes (6.7 hour) per day and and observable time is 446 minutes (7.4 hour) per day. Table~\ref{Tab:mon-rad} lists the monthly statistics of solar premium time and observable time, maximum irradiance, altitude and date of four sites. We can see  that the higher the altitude, the higher the maximum irradiance.   At Mt. Saishiteng, the  maximum of solar irradiance is 1103 W/$m^{2}$ and  the  maximum of solar irradiance at Ali reached 1173 W/$m^{2}$. It is easy to understand that the atmosphere transparency of Ali  is the highest since its altitude is the highest (5100 m) among the all sites.  One reason why the observable hour in Nanshan station is too low maybe that we set some part of abnormal irradiance data which appear 1999 W/$m^{2}$  to zero.

\section{Precipitable water vapor}
\label{sect:pwv}
\begin{figure}
   \centering
   \includegraphics[width=13cm, angle=0]{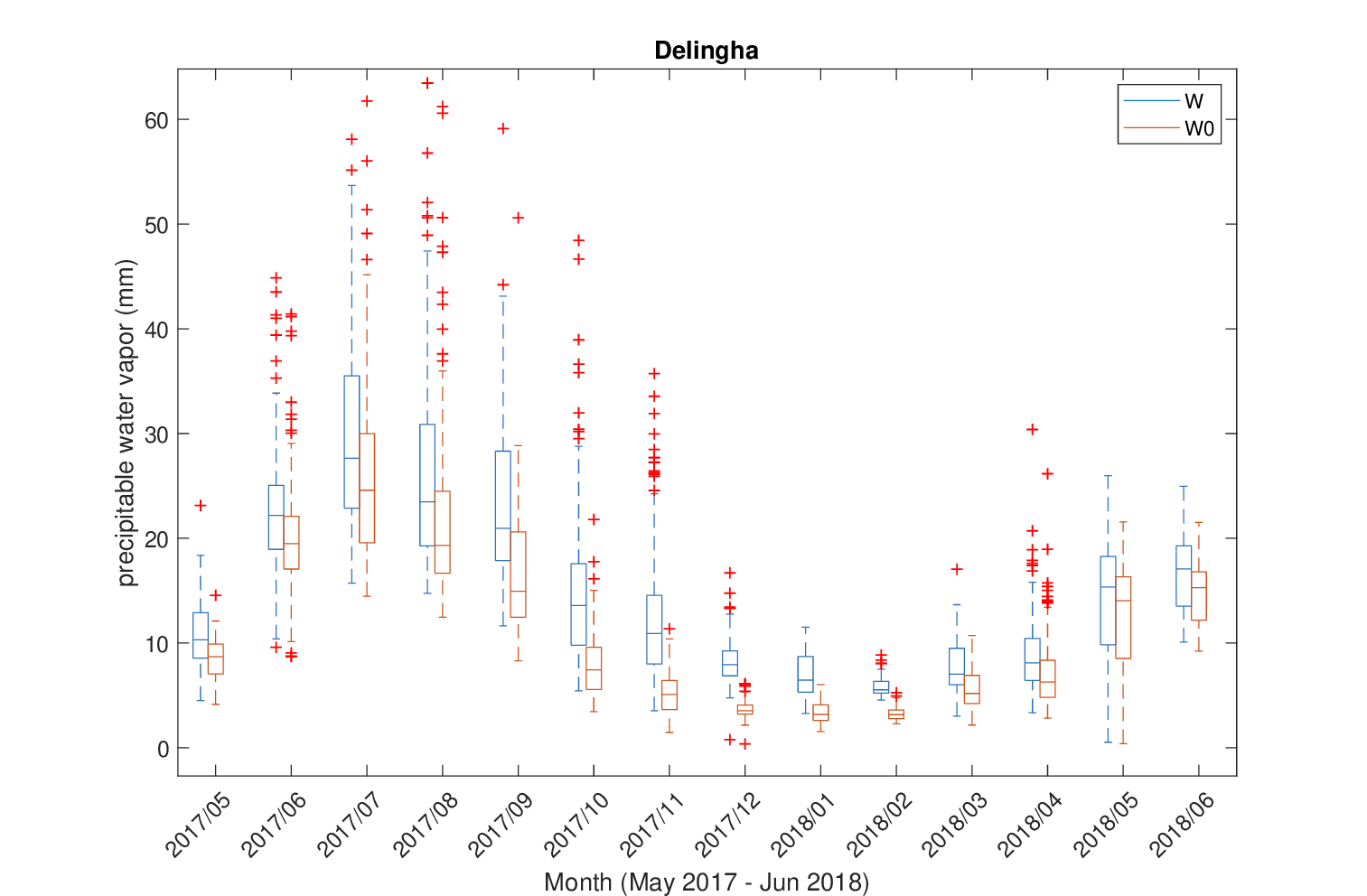}
   \caption{Monthly  variation of precipitable  water vapor in Delingha in Qinghai province. W (blue) represent observational PWV and $W_{0}$ (orange) represent PWV corrected to the zenith.}
   \label{Fig:dewv}
   \end{figure}

\begin{figure}
   \centering
   \includegraphics[width=13cm, angle=0]{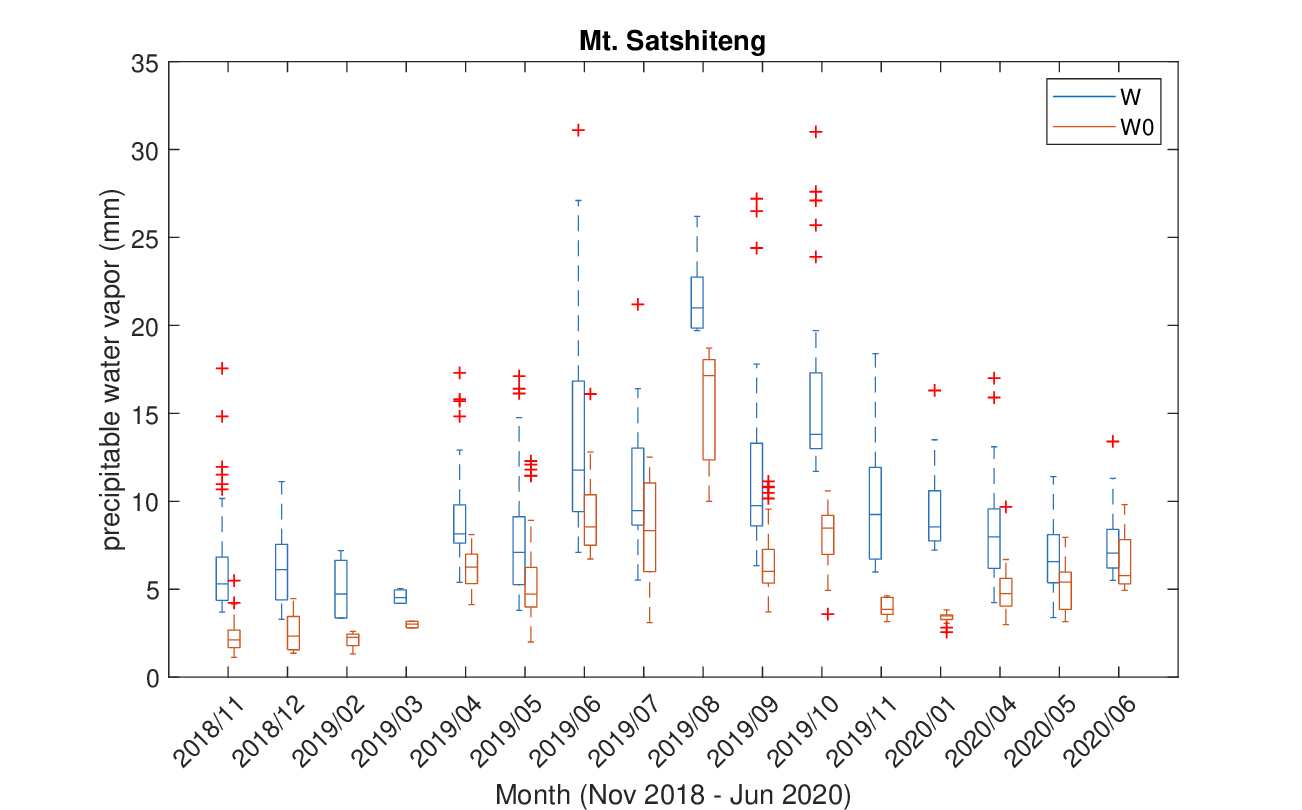}
   \caption{Monthly  variation of precipitable water vapor W (blue) at Mt. Saishiteng during Nov. 18, 2018-July 06, 2020.  W$_{0}$ (orange) indicates the value W with correction to the zenith ( precipitable water vapor divided by air mass). The upper tip, upper top of a box, mid-bar in a box, bottom of a box and lower tip represent 95\%, 75\%, 50\%, 25\% and 5\% of the measured data for each month respectively.}
   \label{Fig:saip}
   \end{figure}

\begin{figure}
   \centering
   \includegraphics[width=10cm, angle=0]{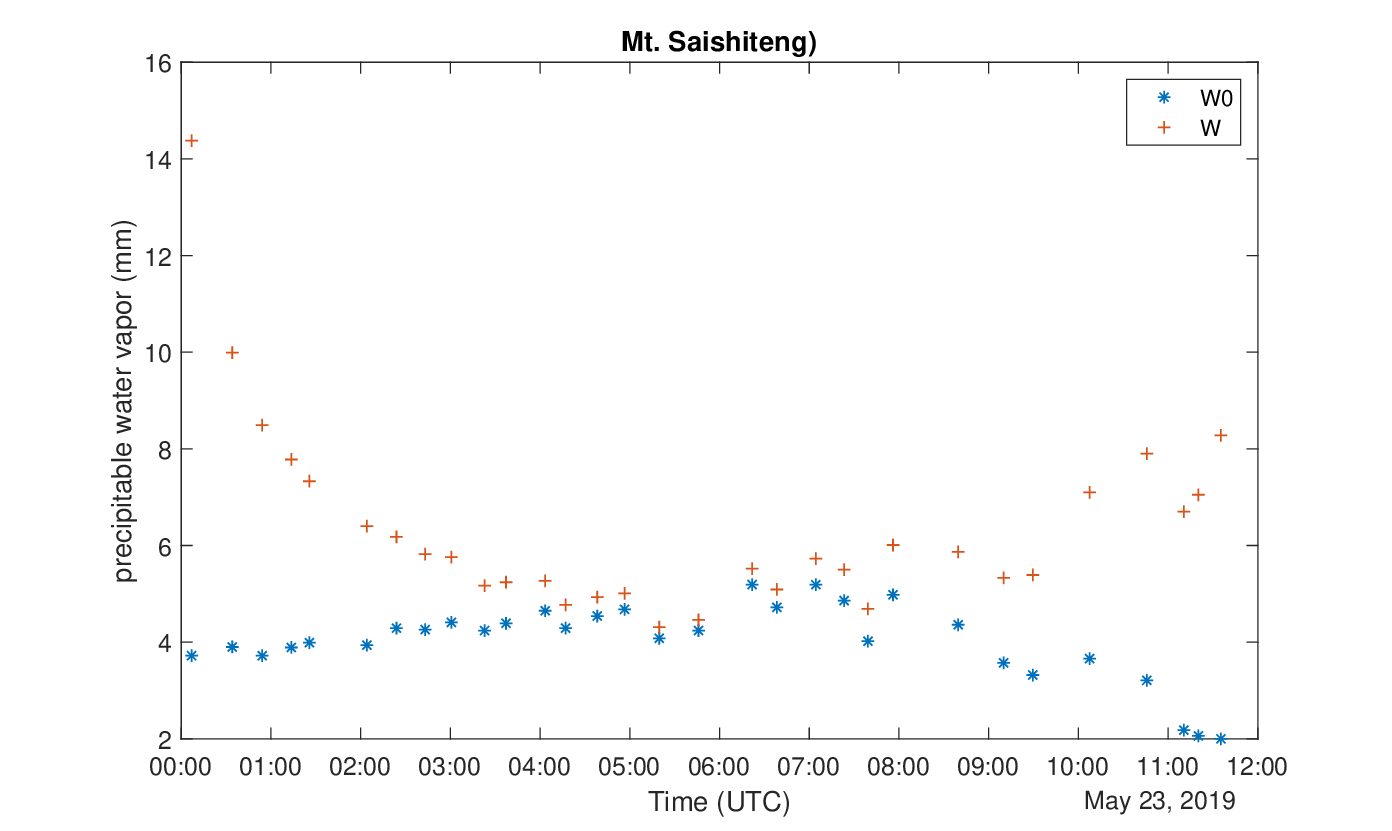}
   \caption{Hourly variation within a day of PWV on May 23, 2019 at Mt. Saishiteng. }
   \label{Fig:saipvd}
   \end{figure}

 The precipitable water vapor (PWV) can be measured by several methods such as radiosonde balloons, radiometers from both ground and satellites, Sun photometers, lunar photometers, GPS receivers, Fourier transform infrared spectrometers and others (\citealt{Qian+etal+2019}).  In the AIMS site testing survey, the precipitable water vapor was measured by a spectrometer which measures residual intensity of the absorption line of  $H_{2}$O molecule centered at 935 nm.  When the tube of spectrometer is pointing to the Sun, precipitable water vapor is calculated from intensities at 935 nm and 889nm from the function below.
  \begin{equation}
 {R} = {0.59}^{\sqrt{W}}%{-\frac{1}{3}}]%{-1/3}(\frac{\lambda}{r_{0}})^{\frac{5}{3}}(\frac{\lambda}{D})^{\frac{1}{3}}%\
\end{equation}
 where R is the residual intensity; It is the ratio between the intensity at 935 nm and 889 nm.  W is  PWV.\\
\\
, The PWV measurementsat Nanshan station and Ali station were obtained in only a few days which is not statistically significant. The observations of precipitable water vapor was carried out every 30 minutes in clear day between May 2017 to June 2018 at Delinagha station.  Its monthly variation of PWV is shown in Figure~\ref{Fig:dewv} with median value W of 11.5 cm and $W_{0}$ of 7.0 cm. W indicates the actual value of PWV measured when the spectrometer pointing to the Sun while $W_{0}$ indicates the corrected value of W to local zenith ---that is W$_{0}$=W/a, where a is the air mass.  Figure~\ref{Fig:saip}  shows monthly variation of precipitable water vapor during the period from Nov. 2018 to June 2020 at Mt. Saishiteng. Its  median value of W  and $W_{0}$ are 8.07 mm and 5.25 mm respectively. In the winter season,  Mt. Saishiteng's  PWV is low with a median value of $W_{0}$ 2.1 mm. In Figure~\ref{Fig:saipvd}, the hourly variation within a day of PWV show that in the early morning, the W value is high due to great optical thickness and the W reached its lowest value after the Sun passes the local meridian.

%               one-column-spanning table
\begin{table}
\begin{center}
\caption[]{ Annual wind speed (m/s) statistics of Ali. Nanshan, Delingha station and  Mt. Saishiteng.}
\label{Tab:annu-winds}

%%Please Capitalize the First Letter of Each Notional Word in table's caption

 \begin{tabular}{clclc}
  \hline\noalign{\smallskip}
site &  average      & median & high &  observation period    \\
  \hline\noalign{\smallskip}
Ali & 5.02 & 4.7     &31.2 &  April 2017-April 2018  \\ % new variable
Nanshan  & 2.05    &   2.24     & 11.6     &    April 2017-May 2018         \\
Delingha  & 1.63     &   1     & 14.7   &    April 2017-May 2018        \\
Mt. Saishiteng  & 3.5 & 3.2     & 26.3  & July 2019-June 2020 \\ % new variable
  \noalign{\smallskip}\hline
\end{tabular}
\end{center}
\end{table}

%________________________________________ Table 2: Use_of_the routines
\begin{table}
\begin{center}
\caption[]{ Monthly average temperature at Mt. Saishiteng.}
\label{Tab:month-temp}

%%Please Capitalize the First Letter of Each Notional Word in table's caption

 \begin{tabular}{clclc}
  \hline\noalign{\smallskip}
month &  average      & median & high &  low    \\
  \hline\noalign{\smallskip}
August & 8.9 & 8.6     & 19.3  & 0.5 \\ % new variable
September  & 4.6     &   4.3     & 15.9     &     -4.4        \\
Octomber  & 3.4     &   4.1     & 8.6   &      -11.6        \\
November  & -8.5 & -8.0     & 0.6  & -18.4 \\ % new variable
January  &-14.5     &   -15.0     & -5.5     &     -20.6        \\
February  & -1.6     &  -11.8     & -3.1   &      -20.5        \\
March & -8.4 & -8.2     & 0  & -15.8 \\ % new variable
April  & -3.8     &   -4     & 7.6     &     -11.2        \\
May  & -0.3     &  -0.7     & 10.8   &      -10.1        \\
  \noalign{\smallskip}\hline
\end{tabular}
\end{center}
\end{table}
%               one-column-spanning table
\begin{table}
\begin{center}
\caption[]{ Monthly statistics of premium time, observable time, solar maximum irradiance, altitude and month of observation of Ali, Nanshan, Delingha station and  Mt. Saishiteng.}
\label{Tab:mon-rad}

%%Please Capitalize the First Letter of Each Notional Word in table's caption

 \begin{tabular}{cccccc}
  \hline\noalign{\smallskip}
site &  premium (hour)      & observable time (hour) & maximum irradiance (W/$m^{2}$) & altitude (m) & month of observation   \\
  \hline\noalign{\smallskip}
Ali & - & -      & 1173 & 5100 & July 2017  \\ % new variable
Nanshan  & 1.95*    &   2.8*     & 995 & 2000  &   June 2017         \\
Delingha  & 4.95     &   6     & 1036 & 3100 &    June 2017        \\
Mt. Saishiteng  & 6.7 & 7.4     & 1103 & 4200 & August 2019 \\ % new variable

  \noalign{\smallskip}\hline
\end{tabular}
\end{center}
*maximum irradiance refers to maximum solar irradiance among all data at one site.
\end{table}
%
%               one-column-spanning table
%________________________________________ Table 2: Use_of_the routines
\begin{table}
\begin{center}
\caption[]{ Parameters and
technical specifications of Mt. Saishiteng S-DIMM.}
\label{Tab:dimm-para}
%%Please Capitalize the First Letter of Each Notional Word in table's caption
 \begin{tabular}{clcl}
  \hline\noalign{\smallskip}
% &                      \\
  \hline\noalign{\smallskip}
telescope and mount\\
model & Celestron CGX 1100   \\ % new variable
diameter & 280                  \\
focal length  & 2800                  \\
mount & ASTROOM ATR Aries                \\
Hartmann Mask\\
sub-aperture diameter & 50                  \\
sub-aperture separation  & 220             \\
deviation angle & 10'                    \\
solar filter & BAADER Film \\
sensor\\
CMOS model & ZWO ASI 174MM                    \\
size  & 11.3*7.1 mm\\
pixels number & 1936$\times$1216                 \\
field of view & 13'$\times$8.1'                \\
one pixel  & 0.44" \\
\noalign{\smallskip}\hline
\end{tabular}
\end{center}
\end{table}

\section{Daytime seeing observation}
\label{sect:seeing}
\begin{figure}
   \centering
   \includegraphics[width=10cm, angle=0]{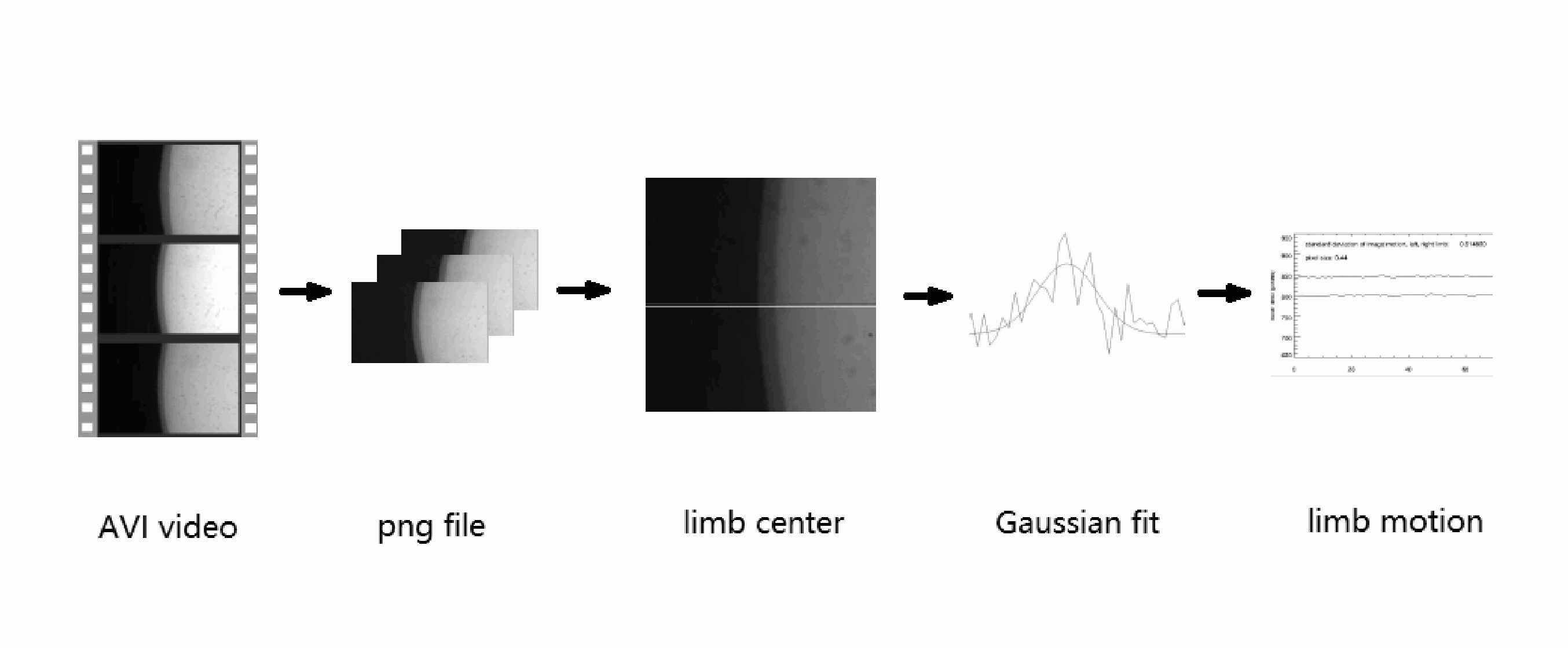}
   \caption{Process of seeing data observation and reduction of S-DIMM at Mt. Saishiteng.}
   \label{Fig:pro2}
   \end{figure}

In Saishiteng Mountain, the S-DIMM is installed on a 10 meter high tower whose pier stood alone separately to the platform ( Figure~\ref{Fig:tower}). It consists of  a Celestron C11 XLT telescope with a clear aperture 280 mm and a F/10 focal ratio ( Figure~\ref{Fig:lhraddimm} bottom). On the cover of C11 telescope, two holes was opened as sub-apertures and was covered by a Baade film with $10^{-5}$ transmission.  The diameter of two sub-apertures is 5 cm and their separation is 22 cm. Two prisms with separate  angle 10 arc seconds were mounted in the each sub-apertures.  A 1/40 neutral filter was installed before the focal plane. The CMOS camera is  ZWO ASI 174MM with a sensor of  1936$\times$1216 pixels. Each pixel correspondents to 0.44 arc seconds (see  Table~\ref{Tab:dimm-para}).\\
\\
When measuring the seeing condition, the solar images were recorded on a AVI video  with an exposure time of 14 ms, then 100 png files were extracted from the AVI video. At first, a intensity distribution along a horizontal direction crossing solar eastern or wastern limb is drawn and then the profile of  intensity gradient of solar eastern or western limb along horizontal direction is a Gaussian curve. The positions of two limb were determined at the center by using Gaussian fit (Figure~\ref{Fig:pro2}). The variation $\sigma$ of  distance between two limb is calculated from the differences of position of two solar limbs.\\

\begin{figure}
   \centering
   \includegraphics[width=12cm, angle=0]{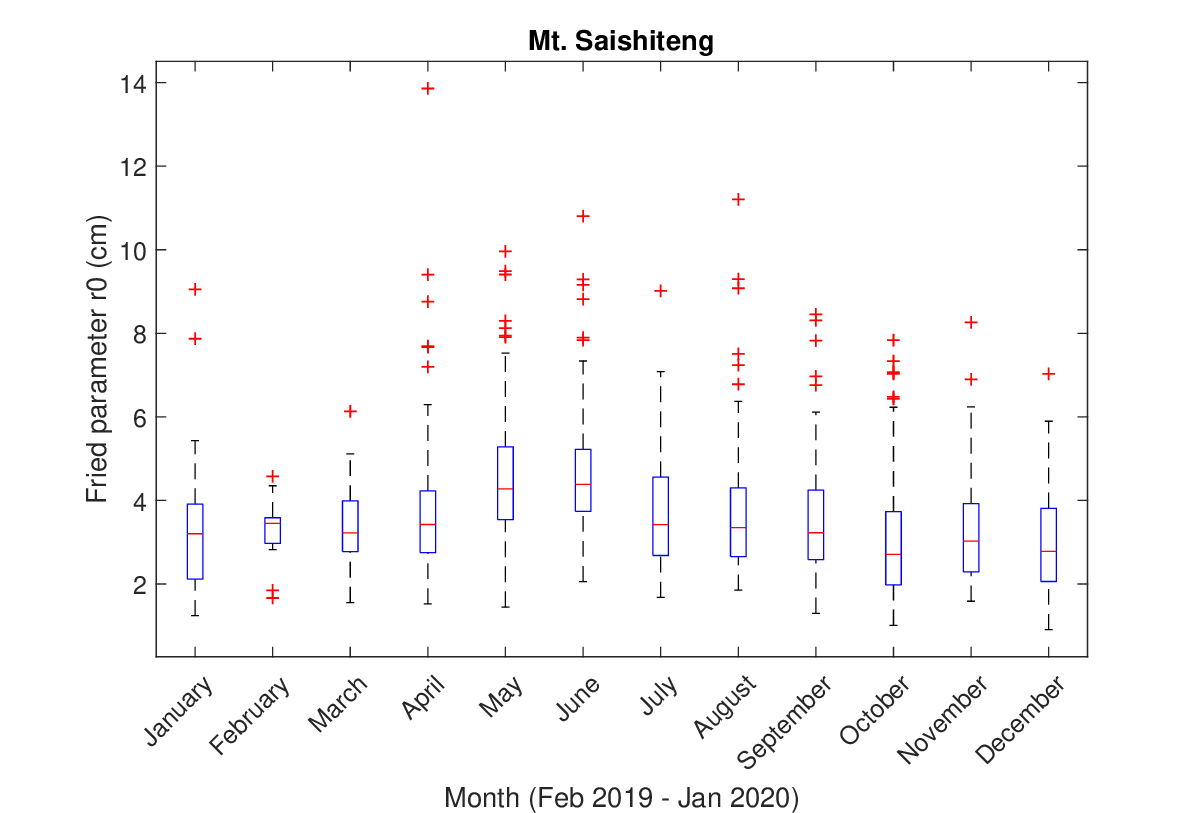}
   \caption{Monthly  variation of daytime seeing $ r_{0}$ at Mt. Saishiteng. The upper tip, upper top of a box, mid-bar in a box, bottom of a box and lower tip represent 95\%, 75\%, 50\%, 25\% and 5\% of the measured data for each month respectively.}
   \label{Fig:seeing}
   \end{figure}

\begin{figure}
   \centering
   \includegraphics[width=12cm, angle=0]{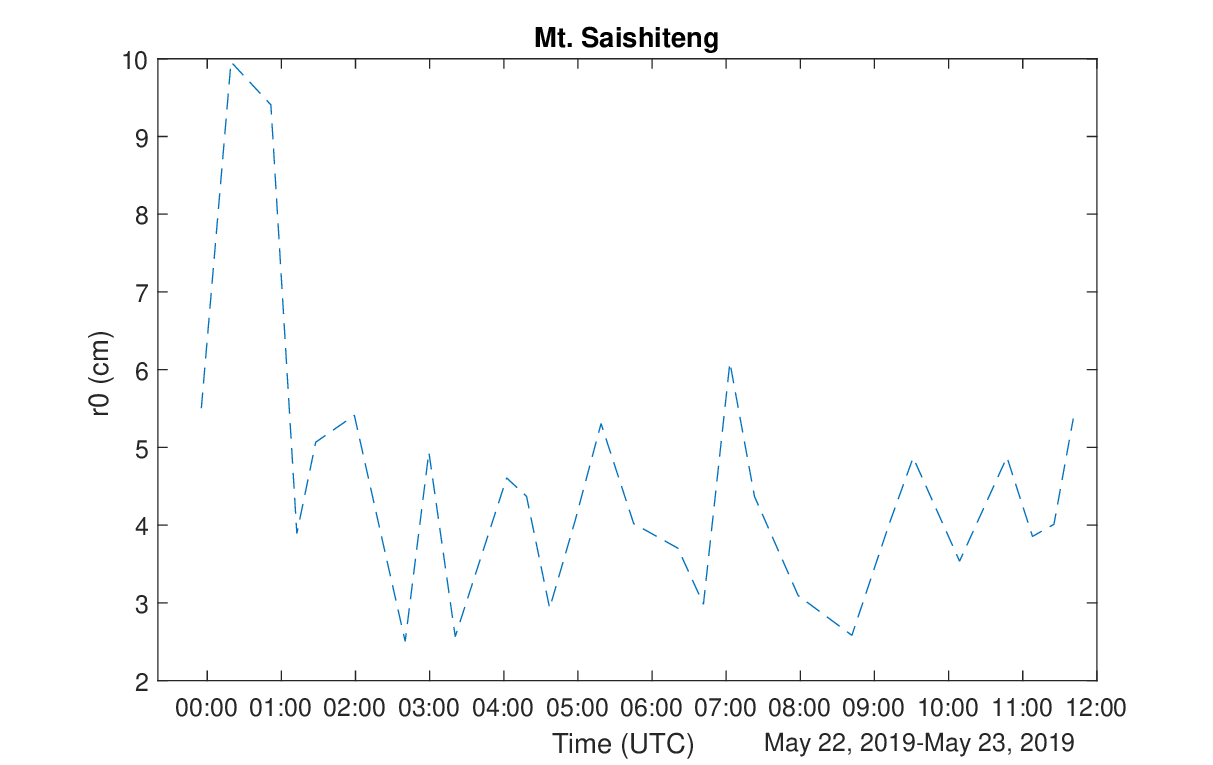}
   \caption{Hourly variation of daytime seeing  condition $ r_{0}$ at Mt. Saishiteng on May 23, 2019.}
   \label{Fig:seep}
   \end{figure}

\begin{figure}
   \centering
   \includegraphics[width=12cm, angle=0]{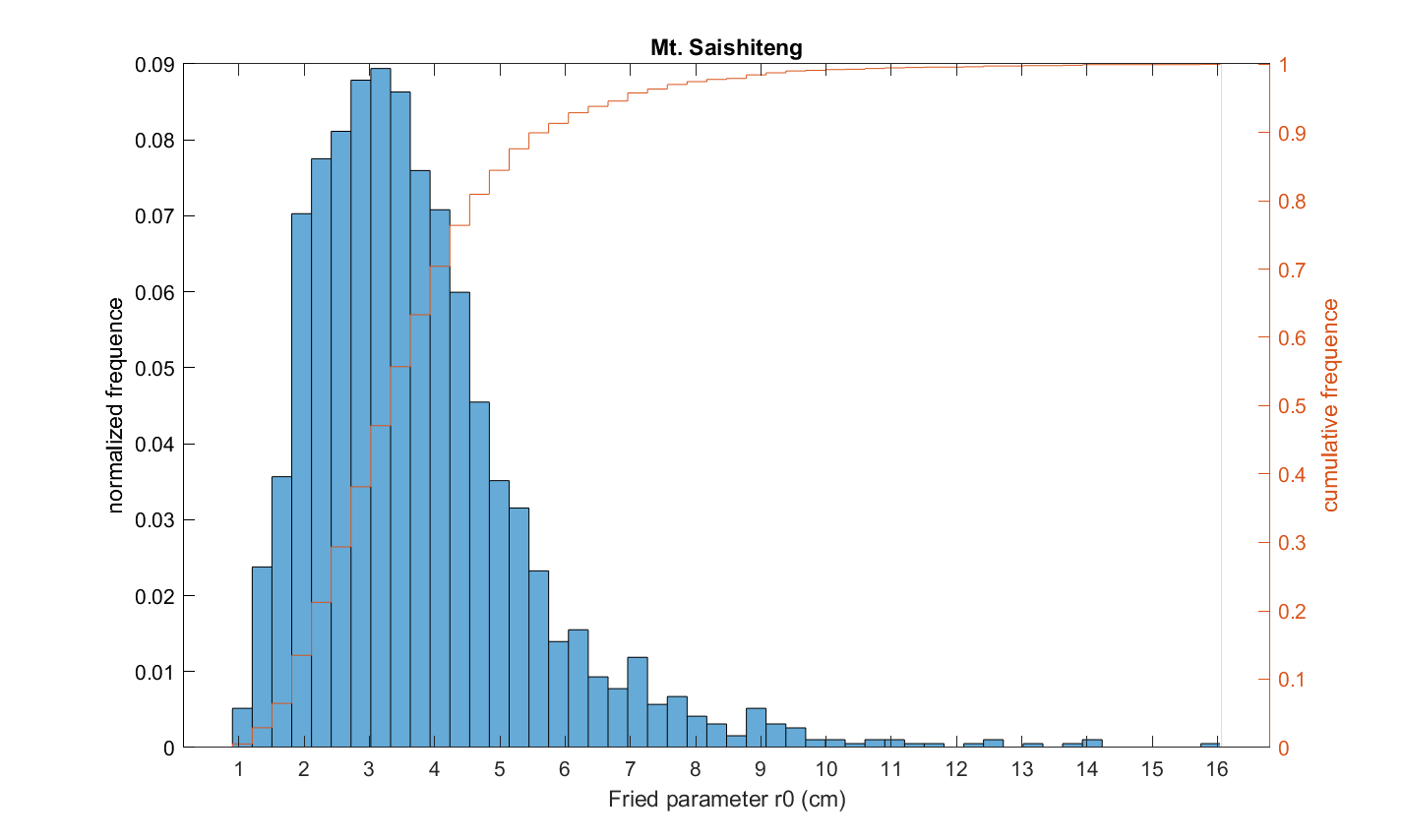}
   \caption{Daytime Fried parameter $ r_{0}$  distributions (blue) and cumulative distribution functions (orange) at Mt. Saishiteng. It peaks at 3.2 cm, 90\% is than 5.6 cm, 70\% is less than 4.3 cm and 30\% is less than 2.7 cm.  }
   \label{Fig:cum}
   \end{figure}

\begin{figure}
   \centering
   \includegraphics[width=14cm, angle=0]{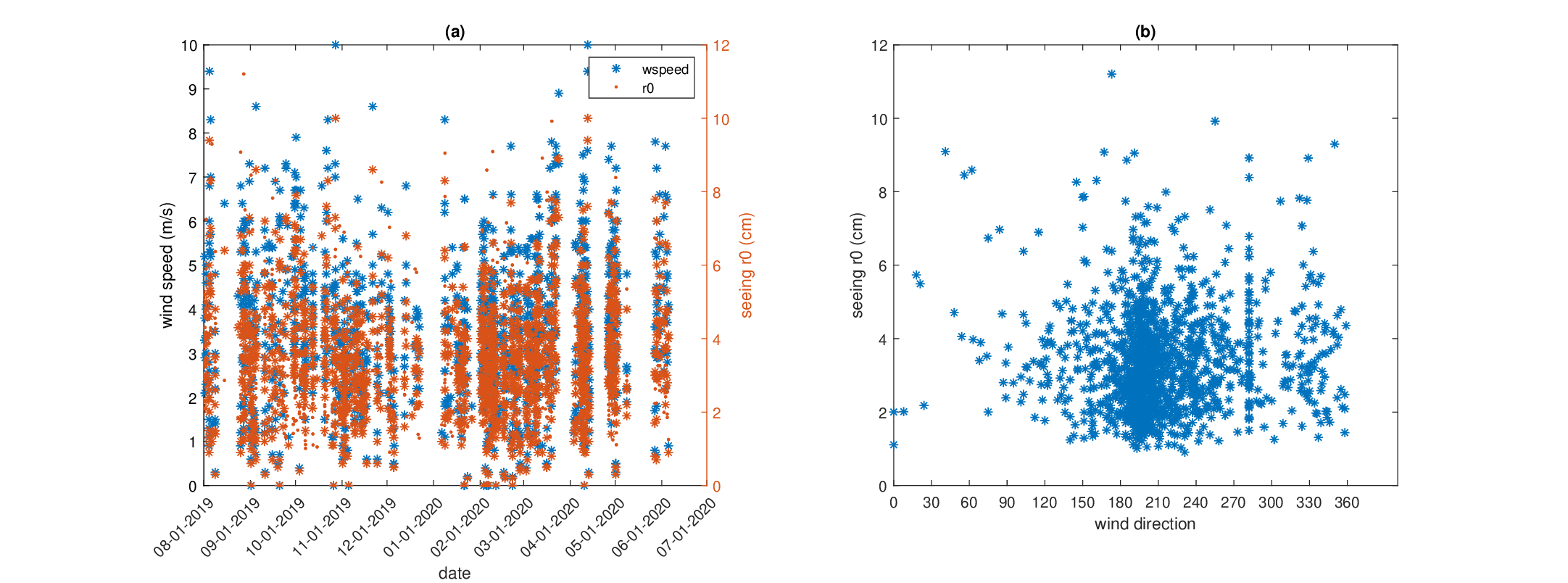}
   \caption{Comparison of daytime Fried parameter $ r_{0}$ (orange) distribution with wind speed in blue color (a), and wind direction (b). 0 degree indicates the north direction. }
   \label{Fig:seew}
   \end{figure}

Then the Fried parameter $r_{0}$ can be calculated from variation ${\sigma}^2$ (\citealt{Sarazin+Roddier+1990}):
\begin{equation}
 {\sigma}^2 = 2{\lambda}^{2}{r_{0}}^{-\frac{5}{3}}[0.179D^{-\frac{1}{3}}-0.0968d^{-\frac{1}{3}}]%{-\frac{1}{3}}]%{-1/3}(\frac{\lambda}{r_{0}})^{\frac{5}{3}}(\frac{\lambda}{D})^{\frac{1}{3}}%\
\end{equation}

or in a simple form (\citealt{Tokovinin+2002})\\

\begin{equation}
 {\sigma}^2 = K (\frac{\lambda}{r_{0}})^{\frac{5}{3}}(\frac{\lambda}{D})^{\frac{1}{3}}%\
\end{equation}

where D is the diameter of sub-aperture, and K is the coefficient:
\begin{equation}
 k = 0.364(1-0.532b^{-\frac{1}{3}}-0.024b^{-\frac{7}{3}})%
\end{equation}

where b=d/D, d is apertures separation. \\
\\
Figure~\ref{Fig:seeing} shows the monthly variation of Fried parameter r$_{0}$ at Mt. Saishiteng with a median value of 3.4 cm. The total 1935 seeing data was taken in every 30 minutes through remote control in each clear day from November 7, 2018 to June 5, 2020  except the days when site testing instruments was in either outage or off-line.   Figure~\ref{Fig:seep} is the hourly variation of Fried parameter on May 23, 2019. It shows that very good seeing condition appears in the  early morning which  indicates that the atmosphere is not disturbed severely by the sunshine.  Figure~\ref{Fig:cum} shows the distribution histogram with cumulative frequency of daytime Fried parameter and indicates that it peaks at 3.2 cm, 90\% is less than 5.6 cm, 70\% is less than 4.3 cm and 30\% is less than 2.7 cm.  In order to explore the relationship between seeing condition r$_{0}$ and wind direction and speed, we plot a distribution of Mt. Saishiteng's seeing condition with wind speed (Figure~\ref{Fig:seew}a) and wind direction (Figure~\ref{Fig:seew}b) during seeing measurement. It shows an almost isotropic distribution of Fried parameter r$_{o}$ indicating little correlation to the direction of wind, nor the wind speed.

\section{Discussion}
\label{sec t:discussion}
\begin{figure}
\centering
 \includegraphics[width=13cm, angle=0]{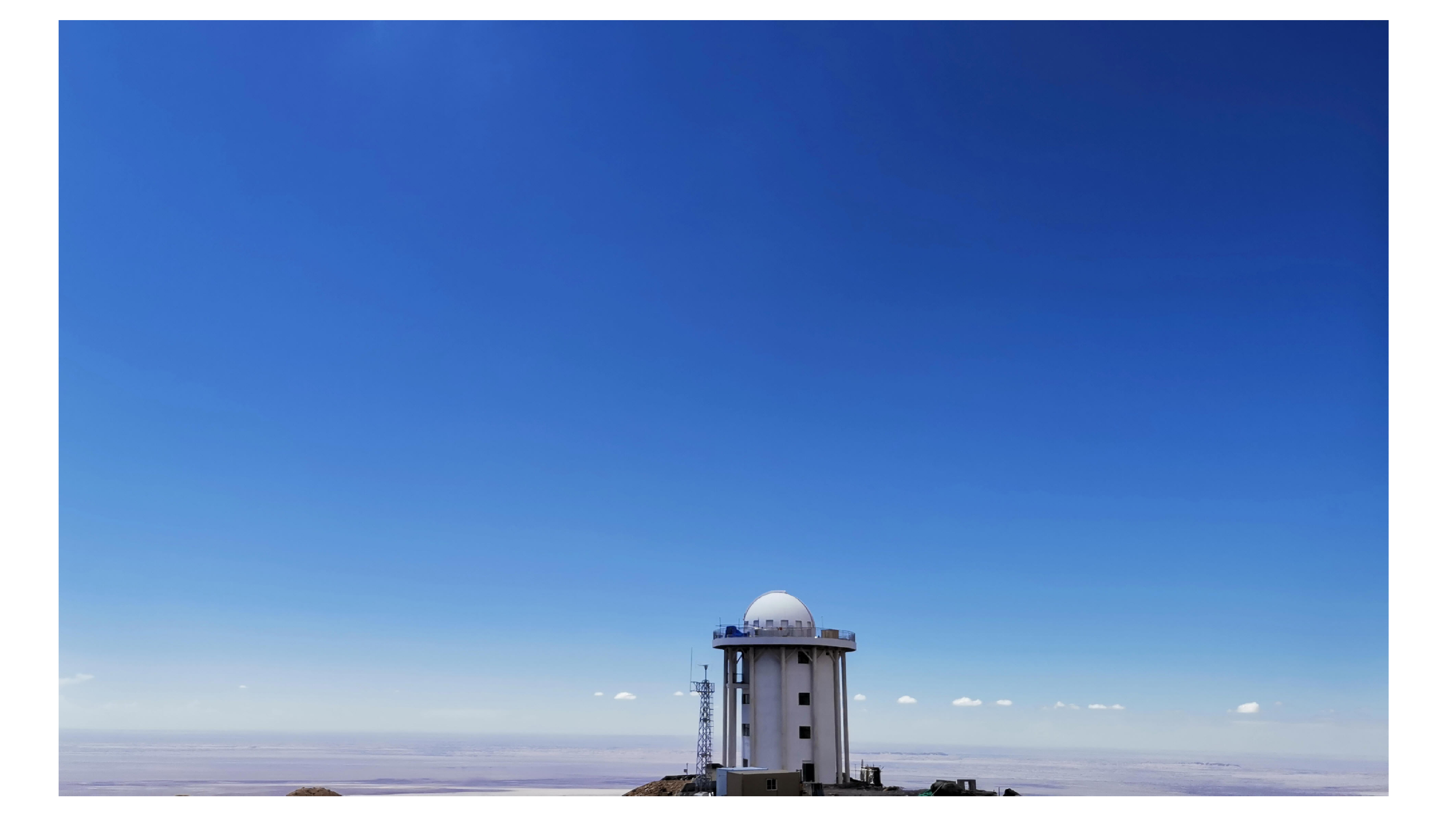}
 \caption{ A wide angle view of AIMS dome on Saishiteng Mountain.}
 \label{Fig:dome}
\end{figure}

For AIMS observation at middle infrared waveband, the PWV is the essential parameter among the all site testing parameters. The Delingha station and Mt. Saishiteng are both located in west part of Qinghai Province with a distance about 400 kilometers, but the the PWV in Delingha is higher than that in Mt. Saishiteng.   One reason behind this is because of their different altitudes (about 1000 m). Another possible reason is that  Delingha is located near the Baiying river's bed. Anyway, our method measuring PWV is not accurate than the method by using a commercial Low Humidity And Temperature Profiling Radiometer (LHATPRO). For example, the median value of PWV is about 0.52 mm at Muztagh-ata in Xinjiang and  2.1 mm at Daocheng in Sichuan (\citealt{Feng+etal+2020}).  The median value of  PWV in European Southern Observatory at Paranel, Chile is 2.5 mm (\citealt{Kerber+etal+2012}).\\
\\
 The average observable time per day in August 2019 is 7.4 hours at Mt. Saishiteng,  which means that there are about 2701 observable hours in a year.  The highest solar radiance of 1103 W/$m^{2}$  indicates the transparency of sky is high at Mt. Saishiteng. Both the living and working condition is tougher in Ali station with altitude of 5100 m above the sea level  and the highest solar radiance of 1173  W/$m^{2}$.  At Delingha station, the average observable time per day is 6 hours in June 2017.\\
\\
For a 1 meter telescope observing the Sun at infra-red band around 12.3 $\mu$m, the median value of Fried parameter $r_{0}$  3.4 cm  at Mt. Saishiteng  is enough and comparable to that in Haleakala, Hawaii for Daniel K. Inoue Solar Telescope (DKIST) with diameter of 4 meters  (\citealt{Ozisik+Ak+2004}).  The median value of Fried parameter r$_{0}$  in Daocheng, Sichuan is 7.2 cm (\citealt{Song+etal+2018}).\\
\\
During initial phase in 2018, the equipments for site testing were carried up to Mt. Saishiteng by a helicopter and we must climb to Mt. Saishiteng on foot in order to install and adjust the telescope. Up to now, an asphalt road, internet and electricity grid have reached to Mt. Saishiteng. Considering the site testing results and other conditions for accommodation and logistics  of all candidate sites comprehensively, the Mt. Saishiteng is selected as the home of AIMS. The actual site of AIMS's dome  was selected at a new hill top with altitude of 4090 meters above sea level (38$^{\circ}$34'26"N, $93^{\circ}$53'45"E)  which is about 900 m south to the original one  due to the possible block of sunshine from a eastern ridge during sunrise in the summer time. By the end of 2021, the construction of AIMS dome has completed and Figure~\ref{Fig:dome} is a wide angle view of the AIMS dome on Mt. Saishiteng taken on August 8, 2022.\\

\begin{acknowledgements}
I would like thank to staffs of Delingha station in Qinghai, Purple Mountain Observatory, Ali station, National Astronomical Observatories and Nanshan station, Xinjiang Astronomical Observatory for their help and  effort during site testing observation.  I would also like to thank Tengfei Song, Yunnan Astronomical Observatory, for preparing optical wedge of S-DIMM.  This work is sponsored by the National Natural Science Foundation of China (NSFC) under the grant numbers of 11427901 and No. 12273059, and National Key R\&D Program of China under grant number of 2021YFA1600500.

\end{acknowledgements}

%      Figure with a new Delingha pwv
%-------------------------------------------------------------
%      A figure as large as the width of the column
   %-------------------------------------------------------------
      %-------------------------------------------------------------
    %-------------------------------------------------------------
   %-------------------------------------------------------------
   %-------------------------------------------------------------
   %-------------------------------------------------------------
  %-------------------------------------------------------------
%————————————————
    %-------------------------------------------------------------
   %-------------------------------------------------------------
 %-------------------------------------------------------------
%-------------------------------------------------------------

%------------------------------------------------------------
%      One column rotated figure
%------------------------------------------------------------
%------------------------------------------------------------
%
%
% one-column-wide figure(occupies half-width of a page)
%  -- This is an old way of graphics inclusion with psfig.sty
%------------------------------------------------------------ Fig1: lightcurve

\label{lastpage}
\end{document}